\documentclass[article]{aa}

\def\msun{M_\odot}
\def\msuny{M_\odot~{\rm y}^{-1}}
\def\msuny9{10^{-9}~\msuny}
\def\mdens{\rm g~cm^{-3}}

\usepackage{color}

\usepackage{graphicx}
\usepackage{natbib}
\usepackage{amssymb,amsmath}
\begin{document}
\title{Keplerian frequency of uniformly rotating neutron
stars and quark stars}
\author{P. Haensel\inst{1}
\and
J.L. Zdunik\inst{1}
\and
M. Bejger\inst{1}
\and
J.M. Lattimer\inst{2}}
 \institute{ N.
Copernicus Astronomical Center, Polish Academy of Sciences,
Bartycka 18, PL-00-716 Warszawa, Poland
{\em jlz@camk.edu.pl} }
 \institute{ N.
Copernicus Astronomical Center, Polish Academy of Sciences,
Bartycka 18, PL-00-716 Warszawa, Poland
\and
Department of Physics and Astronomy, State University of New
York at Stony Brook, Stony Brook, NY 11794-3800, USA\\
{\tt haensel@camk.edu.pl, jlz@camk@edu.pl, bejger@camk.edu.pl,
lattimer@astro.sunysb.edu}}
\date{Received xxx Accepted xxx}
\abstract{}{We calculate Keplerian (mass shedding) configurations of
rigidly rotating neutron stars and quark stars with crusts. We check
the validity of empirical formula for Keplerian frequency, $f_{\rm
K}$,
 proposed by Lattimer \& Prakash,
 $f_K(M)=C\; (M/M_\odot)^{1/2}(R/10~{\rm km})^{-3/2}$~,
 where $M$ is the
(gravitational) mass of Keplerian configuration, $R$ is the
(circumferential) radius of the non-rotating configuration of the same
gravitational mass, and $C=1.04~$kHz.}
{Numerical calculations are performed using precise 2-D
codes based on the multi-domain spectral methods.
We use a representative set of equations of state (EOSs)
 of neutron stars and quark stars.}
{We show that the empirical formula for $f_K(M)$
 holds within a few percent for neutron stars
 with realistic EOSs, provided $0.5\; \msun <M<0.9\; M_{\rm max}^{\rm stat}$,
 where $M_{\rm max}^{\rm stat}$ is the maximum allowable mass of
 non-rotating neutron stars for an EOS, and $C=C_{\rm NS}=1.08~$kHz.
Similar precision is obtained for  quark stars with
$0.5\; \msun <M<0.9\; M_{\rm max}^{\rm stat}$. For maximal crust
masses we obtain $C_{\rm QS}=1.15~$kHz, and the value of
$C_{\rm QS}$ is not very sensitive to the crust mass.
 All our $C$'s are significantly larger
than the analytic value from the relativistic Roche model,
$C_{\rm Roche}=1.00~$kHz. For $0.5\; \msun <M<0.9\; M_{\rm max}^{\rm stat}$,
 the equatorial radius of Keplerian configuration of mass $M$, $R_{\rm K}(M)$,
is,  to a very good approximation,  proportional  to the radius of the non-rotating
star of the same mass, $R_{\rm K}(M)=a\;R(M)$, with $a_{\rm NS}\approx a_{\rm QS}
\approx 1.44$. The value of $a_{\rm QS}$ is very weakly dependent on the
mass of the crust of the quark star. Both $a$'s are smaller than the analytic
value $a_{\rm Roche}=1.5$ from the relativistic Roche model.}{}
\titlerunning{Keplerian frequency of neutron stars}

\maketitle
%
\section{Introduction}
\label{sect:introd}
%

Because of their strong gravity, neutron stars can be very
rapid rotators. In view of the high stability of pulsar frequency
(even the giant glitches produce relatively small fractional
change of rotation frequency, $\la 10^{-5}$), one can treat
pulsar rotation as rigid. The frequency $f$
of stable rotation of a star of gravitational
mass $M$ and baryon mass lower than the maximum
allowable for non-rotating
stars is limited by the (Keplerian)
 frequency $f_{\rm K}$ of a test particle co-rotating
on an orbit at the stellar equator.  The relation
between $f_{\rm K}$ and stellar gravitational mass $M$,
$f_{\rm K}=f_{\rm K}(M)$,
depends on the (unknown) equation of state (EOS) at
supranuclear densities. Both quantities, $M$ and $f$, are
measurable, and the condition implied by a measured frequency $f_{\rm
obs}$ of a pulsar of mass $M$, $f_{\rm obs}<f_{\rm K}(M)$, could be
used to constrain theoretical models of dense matter
(for a recent review of theory of dense matter see \citealt{NSB1}).
Numerical calculation of $f_{\rm K}(M)$  requires precise,
time consuming 2-D calculations of stationary rotating
configurations in general relativity. Therefore, the search
for a sufficiently precise approximate but universal
formula for $f_{\rm K}(M)$  is of great interest. \cite{LP2004}
proposed an approximate  empirical formula
$f_K(M)\approx
C(M/M_\odot)^{1/2}(R/10~{\rm km})^{-3/2}$, where $R=R(M)$ is
the circumferential radius of static star of mass $M$, and
$C=1.04~$kHz does not depend on the EOS. In the present paper we
calculate precise 2-D models of rapidly rotating
neutron stars  and quark stars with different EOSs.  We use
 the relativistic Roche model \citep{STW1983} to motivate the empirical
formula for $f_{\rm K}(M)$ proposed by
\cite{LP2004}.  We calculate the optimal value of prefactor $C$ and we
establish limits to the validity of the empirical formula.

As of this writing, the maximum rotation frequency of a pulsar is
716 Hz (PSR J1748$-$2446ad, \citealt{Hessels2006}).
\citet{Kaaret2007} reported a discovery of oscillation
frequency 1122 Hz in an X-ray burst from the X-ray transient,
XTE J1739-285, and concluded that "this oscillation frequency
suggests that  XTE J1739-285 contains the fastest rotating
neutron star yet found", but this observation has not been
confirmed or reproduced.

The problem of the constraint $f^{\rm EOS}_{\rm max}(M)>f_{\rm
obs}$ was already considered by \citet{STW1983}
after the epochal discovery of the first millisecond pulsar,
PSR 1937+214 with $f_{\rm obs}=641 {\rm Hz}$ \citep{BackerKH1982}.
\citet{STW1983} used a
formula for $f_{\rm max}(M)$ based on the relativistic
Roche model. After the announcement of ill-fated discovery of a 2 kHz
pulsar, this formula was  used to show that nearly all EOSs of
dense matter existing at that time were ruled out by this
observation \citep{STW1989}.

In Sect.\ \ref{sect:Roche} we summarize results obtained with the
relativistic Roche model. Sect.\ \ref{sect:realisticEOSs} contains
description of realistic EOSs of nuclear matter, whereas Sect.\
\ref{sect:calculations} provides the assumptions and methods used to
calculate rotating stellar models. In Sect.\ \ref{sect:fit-hadronic} we
check the validity of empirical formula against results of precise 2-D
calculations for ten realistic EOS for neutron stars. Hypothetical
self-bound quark stars with normal crust  are considered in Sect.\
\ref{sect:quarkEOSs}. Sect.\ \ref{sect:MR.stat.rot.} presents the static
and rotating configurations in the mass-radius plane. In Sect.
\ref{sect:Rmax-R} we derive approximate relations between the
circumferential radius of a static configuration and that of a Keplerian
configuration of the same gravitational mass, for neutron stars and
quark stars.  Discussion of our results is presented in Sect.\
\ref{sect:discuss.concl}.
\section{Relativistic Roche model} \label{sect:Roche} 
There exists an instructive model of neutron stars for which an analytic
formula for $f_{\rm K}(M)$ can be obtained (\citealt{STW1983,STW1989}).
It is  a relativistic Roche model, in which the mass of the star is
assumed to be strongly centrally condensed. Consider a continuous
sequence of stationary configurations of constant gravitational
mass $M$,
and rotation frequencies ranging from zero to $f_{\rm K}$. Let the
circumferential radius of non-rotating configuration be $R$. Under the
assumption of an extreme central mass condensation,
\citet{STW1983,STW1989} found an equation satisfied by the coordinates
of the stellar surface (Eq.\ (2) of \citealt{STW1989}). In the special
case of the stellar equator, this equation implies that for normal
equilibrium configurations rotating uniformly at $f$,  the equatorial
circumferential radius $R_{\rm eq}$ satisfies
\begin{equation}
{2GM\over R_{\rm eq}}+(4\pi f)^2 R_{\rm eq}^2={2GM\over R}~,
\label{eq:metric_eq}
\end{equation}
where the extreme central concentration of matter implies the
gravitational mass can be treated as constant and
equal to the static value $M$. The left-hand-side of Eq.\
(\ref{eq:metric_eq}) reaches a minimum at $R_{\rm
eq}=(GM/4\pi^2f^2)^{1/3}$. For the solution to exist for a given $f$,
the value of the left-hand-side at this minimum should not exceed
$2GM/R$, which implies a condition on $f$ \citep{STW1983,STW1989},
\begin{equation}
f\le f_{\rm K}= {1\over 2\pi} \left({2\over
3}\right)^{3/2} \left({GM\over R^3}\right)^{1/2}~. \label{eq:f_fK}
\end{equation}
Therefore, the Keplerian frequency is
\begin{equation} f_{\rm K}^{\rm Roche}(M)= 1.00~{\rm
kHz}\left({M\over M_\odot}\right)^{1/2} \left({R\over 10~{\rm
km}}\right)^{-3/2}~. \label{eq:Roche2} \end{equation}
 As stated in \cite{STW1989}, ``the Relativistic Roche model provides a
surprisingly accurate estimate of the maximum rotation rate
along constant-rest mass sequences '' for many realistic EOSs.

It is easy to show that  an additional relation between
Keplerian and static configuration
can be obtained.  Namely, using Eq.\ (\ref{eq:metric_eq}),
one obtains
a  formula expressing $R_{\rm eq}$ for Keplerian configuration, $R_{\rm K}$,
in terms of $R$ for static configuration of the same mass
$M$,
\begin{equation}
R_{\rm K}(M)={3\over 2}R(M)~.
\label{eq:R_K.Roche}
\end{equation}

The formula for $f_{\rm K}(M)$, Eq.\ (\ref{eq:f_fK}), and that
for $R_{\rm K}(M)$,  imply that $f_{\rm K}(M)$ is equal to the
orbital frequency of a test particle orbiting at $r=R_{\rm K}$ in
the Schwarzschild space-time around a point mass $M$ at $r=0$.
An {\it approximate} equality
$f_{\rm K}(M)\approx f_{\rm orb}^{\rm Schw.}(M,R_{\rm K})$ was
shown to be valid within a few percent
 for normal neutron stars and quark stars \citep{Bejger1122Hz}.
This relation
 {\it holds  strictly} for the relativistic Roche model,
\begin{equation}
f_{\rm K}^{\rm Roche}(M)=f^{\rm Schw.}_{\rm orb}(M,R_{\rm K})=
{1\over 2\pi}\left({GM\over R_{\rm K}^3}\right)^{1/2}~.
\label{eq:Schw.Roche}
\end{equation}
%
\section{Realistic EOSs of hadronic matter}
\label{sect:realisticEOSs}
%
\begin{table}
\caption{Equations of state of neutron star core. N - nucleons and
leptons. NH - nucleons, hyperons, and leptons. Exotic states of
hadronic matter are indicated explicitly. Maximum allowable mass
for  non-rotating stars, $M_{\rm max}^{\rm stat}$,
and the circumferential radius of
non-rotating stars of $1.4~M_\odot$, $R_{1.4}$, are
given in last two columns, respectively.}
\begin{center}
\begin{tabular}[t]{|c|c|c|c|c|}
\hline \raisebox{-1.5ex}{EOS} & \raisebox{-1.5ex}{model} &
\raisebox{-1.5ex}{ref.}
& $M_{\rm max}^{\rm stat}$& $R_{1.4}$\\[-1.ex]
&&&[$M_\odot$]&[km]\\\hline\hline \raisebox{-1.5ex}{FPS} & N,
energy& \raisebox{-1.5ex}a&
\raisebox{-1.5ex}{1.800}&\raisebox{-1.5ex}{10.85}\\[-.5ex]
&density functional&&&\\\hline \raisebox{-1.5ex}{GN3}&N,
relativistic& \raisebox{-1.5ex}b&
\raisebox{-1.5ex}{2.134}&\raisebox{-1.5ex}{14.22}\\[-.5ex]
&mean field&&&\\\hline
 \raisebox{-1.5ex}{DH}& N, energy & \raisebox{-1.5ex}c&
\raisebox{-1.5ex}{2.048}&\raisebox{-1.5ex}{11.69}\\[-.5ex]
&density functional&&&\\\hline \raisebox{-1.5ex}{WFF1}& N,
variational&\raisebox{-1.5ex}d&
\raisebox{-1.5ex}{2.136}&\raisebox{-1.5ex}{10.47}\\[-.5ex]
&theory&&&\\\hline \raisebox{-1.5ex}{APR}& N, variational&
\raisebox{-1.5ex}{${\rm d}^\prime$}&
\raisebox{-1.5ex}{2.212}&  \raisebox{-1.5ex}{11.42}\\[-.5ex]
&theory&&&\\\hline \raisebox{-1.5ex}{BGN1H1} & NH, energy&
\raisebox{-1.5ex}e&
\raisebox{-1.5ex}{1.630}&  \raisebox{-1.5ex}{12.90} \\[-.5ex]
&density functional&&&\\\hline \raisebox{-.5ex}{BBB} &
\raisebox{-.5ex}{N, Brueckner theory}&\raisebox{-.5ex}f&
\raisebox{-.5ex}{1.920}&\raisebox{-.5ex}{11.13}\\\hline
\raisebox{-1.5ex}{GNH3} & NH,  relativistic &  \raisebox{-1.5ex}g&
\raisebox{-1.5ex}{1.964}&  \raisebox{-1.5ex}{14.20}\\[-.5ex]
&mean field &&&\\\hline \raisebox{-1.5ex}{GMGS-Km} & N + mixed &
\raisebox{-1.5ex}h&
\raisebox{-1.5ex}{1.422}&  \raisebox{-1.5ex}{9.95}\\[-.5ex]
&N-kaon condensed&&&\\\hline \raisebox{-1.5ex}{GMGS-Kp} & N + pure &
\raisebox{-1.5ex}i &
\raisebox{-1.5ex}{1.420}&\raisebox{-1.5ex}{13.20}\\[-1.ex]
&kaon condensed&&&\\
 \hline
\end{tabular}
\end{center}
{References for the EOS:
 a -  \cite{ FPS1989};
 b -  \cite{Glend1985}; c - \cite{DH2001};
 d -  A14 Argonne NN potential and Urbana VII three body
 NNN potential, from \cite{WFF1988};
 ${\rm d}^\prime$  -  A18  Argonne NN potential with
 relativistic corrections and Urbana  modified UIX$^*$
 NNN potential model,
  from  \cite{APR1998};
 e -   \cite{BG1997};
 f - Paris two-body NN potential and Urbana
 UIX three body NNN potential, from \cite{BBB1997};
 g - \cite{Glend1985};
  h, i - kaon condensate models with $U_K^{\rm lin}=-130~$MeV,
  \cite{PonsKcond2000}
 }
 \label{tab:EOSs-had}
\end{table}

 In view of a high degree of our ignorance concerning
 the  EOS of dense
 hadronic matter at supranuclear densities ($\rho>3\times 10^{14}~
 {\rm g~cm^{-3}}$),  it is common to consider a set of EOSs based
 on different dense matter theories (for a review, see \citealt{NSB1}).
We used ten  theoretical EOSs.
These  EOSs are listed in Tab.\ \ref{tab:EOSs-had},
where the basic informations (label of an
EOS, theory of dense matter, reference to the original paper)
are also collected.

Six  EOSs are based on realistic models involving only
nucleons (FPS, BBB, DH, WFF1,  APR, GN3). Four   remaining
EOSs  are softened at high  density either by the appearance of hyperons
(GNH3, BGN1H1), or a phase transition to
 a kaon-condensed state (GMGS-Km, GMGS-Kp).

For GMGS-Km and  GMGS-Kp models, the hadronic Lagrangian
is the same. However, to get GMGS-Kp one assumes that
the phase transition takes place
between two pure phases and is accompanied by a density jump,
calculated using the Maxwell construction. The  GMGS-Km EOS
is  obtained  assuming that the transition occurs via a mixed
state of two phases (Gibbs construction). A mixed state is
energetically preferred  when the surface tension between the
two phases is below a certain critical value. As the value of
the surface tension is very uncertain, we considered both
cases.

In all EOSs models (except FPS), the core EOS was joined with the
DH EOS of the crust \citep{DH2001}. For the FPS model, the FPS EOS
of the core was supplemented with the FPS crust EOS of
\cite{Lorenz1993}.

Our set of EOSs includes very different types of models. This
is reflected by a large scatter of the maximum allowable masses
of non-rotating stars,
$1.42M_\odot\la M_{\rm max}\la 2.21 M_\odot$, and a range of
circumferential radii of non-rotating stars with $M=1.4 M_\odot$,
$9.95 {\rm km}\la R_{1.4}\la 14.22 {\rm km}$ (Table 1).

\begin{figure}[h]
\centering
\resizebox{3.5in}{!}{\includegraphics[angle=0]{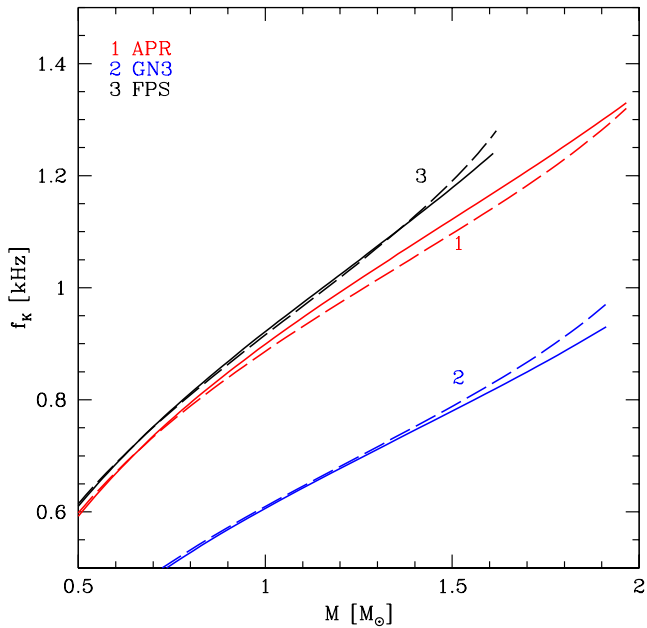}}
\caption{(Color online) Precise values of Keplerian frequency
$f_{\rm K}$ (solid line)
 and those calculated using Eq.\ (\ref{eq:fmax.form}) (dashed line),
 assuming $C=C_{\rm NS}=1.08~$kHz,
 versus stellar mass $M$.
We consider masses $0.5\;\msun <M<0.9\;M_{\rm max}^{\rm stat}$
and APR, GN3, and FPS EOSs. }
\label{fig:fmax.Nucl1}
\end{figure}
\begin{figure}[h]
\centering
\resizebox{3.5in}{!}{\includegraphics[angle=0]{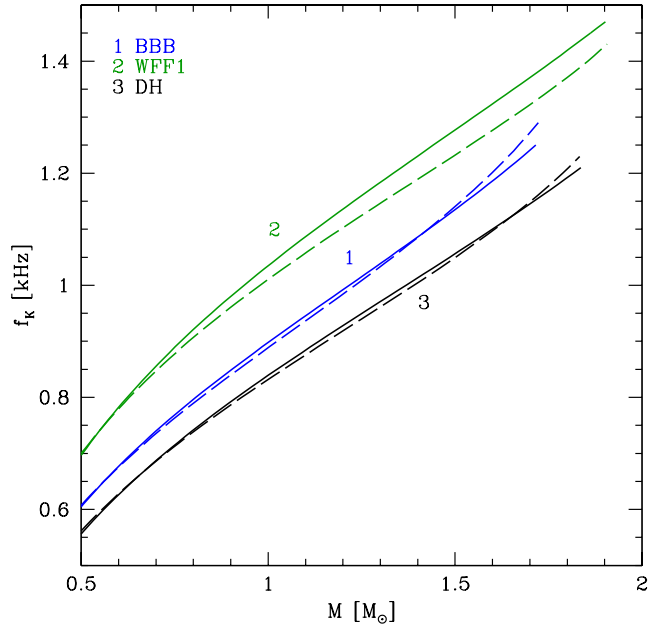}}
\caption{(Color online) Same as Fig.\ \ref{fig:fmax.Nucl1} but
for BBB, WFF1, and DH EOSs.} \label{fig:fmax.Nucl2}
\end{figure}
\begin{figure}[h]
\centering
\resizebox{3.5in}{!}{\includegraphics[angle=0]{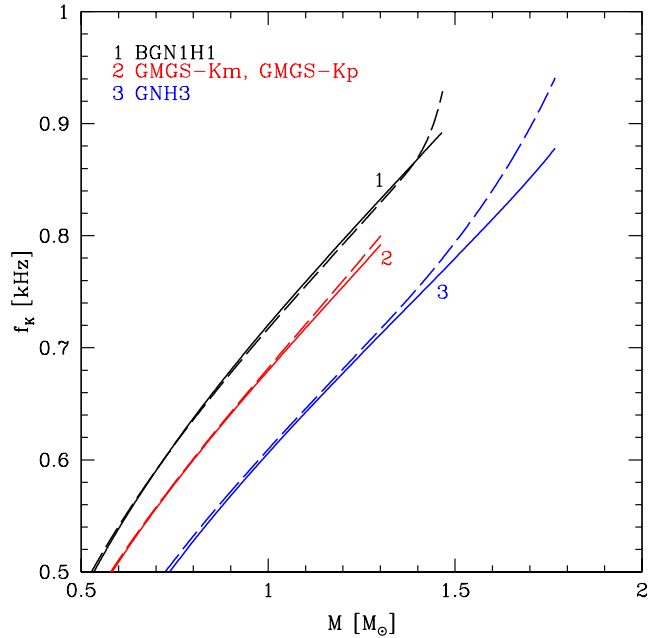}}
\caption{(Color online) Same as Fig.\ \ref{fig:fmax.Nucl1} but
for BGN1H1, GNH3, GMGS-Km and GMGS-Kp EOSs.
Notice that due to a very strong softening by the
kaon condensate and simultaneous constraint $M<0.9\;M_{\rm max}^{\rm stat}$,
the GMGS-Km and GMGS-Kp curves do not contain kaon-condensed segments.
Therefore, the curves for both these EOSs coincide.
 }
\label{fig:fmax.Exo-Hyp}
\end{figure}
\begin{figure}[h]
\centering
\resizebox{3.5in}{!}{\includegraphics[angle=0]{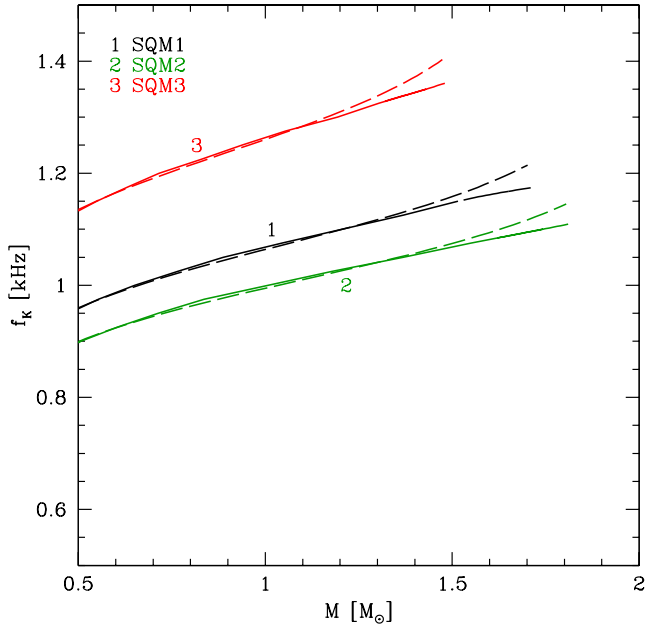}}
\caption{(Color online) Same as Fig.\ \ref{fig:fmax.Nucl1} but
for three EOSs of quark stars. Quarks stars possess maximal
crust, with bottom density $\rho_{\rm b}=\rho_{\rm ND}$.
To get dash lines, we used Eq.\ (\ref{eq:fmax.form}) with
$C=C_{\rm QS}=1.15$.}
\label{fig:fmax.QScrust}
\end{figure}
%
\section{Calculating stationary rotating configurations }
\label{sect:calculations}
%

The  stationary configurations of rigidly rotating neutron
stars have been computed in full general relativity by solving
the Einstein equations for stationary axisymmetric spacetime
(see \citealt{BGSM1993,GourgSS1999} for the
complete set of partial
differential equations to be integrated). The numerical
computations have been performed  using  the ${\tt rotstar}$
 code from the LORENE library (${\tt
http://www.lorene.obspm.fr}$). The code implements a
multi-domain spectral method introduced in  \cite{BGM1998}. A
description of the code can be found in \cite{GourgSS1999}. The
accuracy of the calculations has been checked by evaluation of
the GRV2 and GRV3 virial error indicators
\citep{GourgoulhonB1994,BonazzolaG1994},
which showed values lower than $\sim 10^{-5}$.

%
\section{Maximum  frequencies for realistic EOSs of neutron stars}
\label{sect:fit-hadronic}
%

In Figs.\ \ref{fig:fmax.Nucl1}$-$\ref{fig:fmax.Exo-Hyp} we compare  precisely
calculated Keplerian frequencies with those
given by the empirical formula
%
\begin{equation}
f_{\rm K}(M)\approx C~\left({M\over
M_\odot}\right)^{1/2}\left({R\over 10~{\rm
km}}\right)^{-3/2}~,
\label{eq:fmax.form}
\end{equation}
%
where $M$ is the gravitational mass of rotating star and $R$
is the radius of the non-rotating star of mass
$M$, $R=R(M)$. The optimal value of the $C$ prefactor is
$C_{\rm NS}=1.08~$kHz. The precision of the empirical formula for
$f_{\rm K}$ stays remarkably high for
$0.5\;\msun<M<0.9M^{\rm stat}_{\rm
max}$. Relative deviations are typically within 2\%, with largest
deviations of at most  $6\%$ for the highest masses.

\begin{figure}[h]
\centering \resizebox{3.5in}{!}{\includegraphics[]{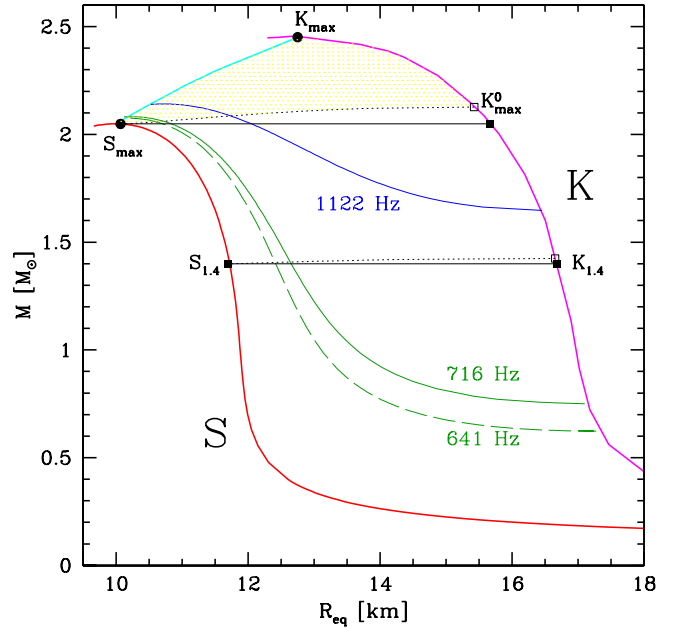}}
\caption{(Color online) Gravitational mass,  $M$, versus
equatorial radius, $R_{\rm eq}$, for static and rigidly
rotating neutron stars, based on the DH EOS. Solid line
$\bf S$: static models ({\it i.e.,} $R(M)$).
Solid line $\bf K$: Keplerian
(mass-shedding) configurations ({\it i.e.,} $R_{\rm K}(M)$).
The area, bounded
by the $\bf S$, $\bf K$ curves and a dash line
${\rm S}_{\rm max}{\rm K}_{\rm max}$, consists of points
 corresponding  to  stationary rotating configurations.
Configurations belonging to a shaded triangular area
 above the dot line ${\rm S}_{\rm max}-{\rm K}^0_{\rm max}$ have baryon mass
 $M_{\rm b}$ larger
 than the maximum allowable baryon mass for non-rotating stars,
 $M^{\rm stat}_{\rm b,max}$.  Three lines corresponding to
  neutron stars rotating stably at $f=641~{\rm Hz},
  716~{\rm Hz}$, and $1122~{\rm Hz}$, are labeled with
  rotation frequencies.
The nearly horizontal dot line ${\rm S}_{1.4}-{\rm K}_{1.4}$
corresponds to configurations with fixed
baryon number equal to that of the non-rotating star of
gravitational mass $M=1.4~M_\odot$.  Generally, solid lines
connecting filled circle with filled square correspond to $M=const.$,
while dot lines connecting filled circle with open
square correspond to stars with $M_{\rm b}=const.$.
 For further explanations
 see the text.}
\label{fig:MRrot}
\end{figure}
\section{Quark stars}
\label{sect:quarkEOSs}
The case of strange stars, built of self-bound quark matter,
is different from that of ordinary neutron stars. Matter
distribution within quark stars has very low density contrast
between the quark core edge and its center. We considered
three EOSs of self-bound quark matter, based on the MIT Bag
Model \citep{FarhiJ1984,Zdunik2000}.
Model parameters are given in Table\
\ref{tab:EOSs-SQM}. Quark stars are likely to have a thin
normal matter crust, with bottom density, $\rho_{\rm b}$,
 not exceeding the neutron drip density $\rho_{\rm ND}\approx
4\times 10^{11}~\mdens$. Maximum mass of normal crusts is
reached for  $\rho_{\rm b}=\rho_{\rm ND}$.

First we consider quark stars with a maximum crust. As we see
in Fig.\;\ref{fig:fmax.QScrust}, precision of empirical
formula within the mass range $0.5~\msun <M<0.9\;M_{\rm max}^{\rm stat}$
is as high as for neutron stars (typical relative deviation
within 2\%, largest deviation of about 4\% at
highest masses). However, the value of $C$ is larger than for
neutron stars, $C_{\rm QS}=1.15$.

Let us consider now quark star models with less massive crusts. These
models were constructed assuming $\rho_{\rm b}<\rho_{\rm ND}$. The effect
on the optimum value $C_{\rm QS}\approx f_{\rm K}(M)\left(M/\msun\right)^{-1/2}
\left(R/10~{\rm km}\right)^{3/2}$ turned out to be very small. At a fixed $M$,
decrease of $\rho_{\rm b}$ leads to an increase of $f_{\rm K}$ (more compact star).
Simultaneously, however, the static value of $R(M)$ decreases, and
therefore both effects cancel out to a large extent. Consequently,
$C_{\rm QS}$ depends rather weakly on the crust mass,
and in principle one may use $C_{\rm QS}=1.15$ for
any crust.
\begin{table}
\caption{Parameters of the bag models for quark stars.
$B$ -  MIT bag constant, $m_s$ - strange quark mass. For
all models the QCD coupling constant equals
$\alpha_{\rm s}=0.2$. Maximum allowable mass
for  non-rotating stars, $M_{\rm max}^{\rm stat}$,
and the circumferential radius of
non-rotating stars of $1.4~M_\odot$, $R_{1.4}$, are
given in last two columns, respectively.}
\begin{center}
\begin{tabular}[t]{|c|c|c|c|c|}
\hline\hline
\raisebox{-1.5ex}{EOS}& $B$ & $m_s c^2 $ & $M_{\rm max}^{\rm stat}$ &
$R_{1.4}$\\[-1.ex]
& [${\rm MeV~fm^{-3}}$] & [${\rm MeV}$]  &[$M_\odot$] & [km]\\
\hline
 SQM1 &    56     &      200   & 1.90 & 11.27\\
 SQM2 &     45    & 185      & 2.02 & 11.86\\
 SQM3 &   67      &    205     & 1.65& 9.94\\
 \hline\hline
\end{tabular}
\end{center}
 \label{tab:EOSs-SQM}
\end{table}

\section{Static and rotating configurations in the mass-radius plane}
\label{sect:MR.stat.rot.}

The formulae for $f_{\rm K}(M)$ are based on a
one-to-one  correspondence between a static configuration S,
belonging to static boundary ${\bf S}$ of the region
of rotating configurations, and the rotation frequency of a  Keplerian
configuration K on the ${\bf K}$ boundary.  This
correspondence is visualized in Fig.\ \ref{fig:MRrot}, based
on the numerical results obtained for the DH EOS.
The frequency of rotation of a Keplerian configuration K is
obtained via the mass and radius of a static configuration $S$
with same $M$. Both configurations are connected by a horizontal
line in the $R_{\rm eq}-M$ plane.

The empirical formula for the {\it absolute upper bound} on $f$ of
stably rotating configurations for a given EOS,
$f_{\rm max}^{\rm EOS}$ \citep{HZ1989,FIP1989,STW1989,LPMY1990,HSB1995},
is of a different character.
It results from an (approximate but precise)
 one-to-one correspondence between the parameters of two
extremal configurations, static ${\rm S}_{\rm max}$
and Keplerian ${\rm K}_{\rm max}$ (filled circles),
and reads
\begin{equation}
f^{\rm EOS}_{\rm max}\approx \mathcal{C} \;
\left({M_{\rm max}^{\rm stat}\over \msun}\right)^{1/2}
\left({R_{M_{\rm max}^{\rm stat}}^{\rm stat}
\over 10\;{\rm km}}\right)^{-3/2}~,
\label{eq:emp.fmax.abs}
\end{equation}
where $\mathcal{C}$ is to a very good approximation independent
of the EOS. We have
$\mathcal{C}_{\rm NS}\approx \mathcal{C}_{\rm QS}=1.22~$kHz
 \citep{HSB1995}. This value is noticeably higher
than $C_{\rm NS}$ or $C_{\rm QS}$, which determine $f_{\rm K}(M)$ for
$0.5 M_\odot <M < 0.9M_{\rm max}^{\rm stat}$.

The functional form of Eq. (\ref{eq:emp.fmax.abs}) is, in fact, exact
in general relativity for uniform rotation of stars with
the so-called minimum period EOS of \cite{KSF1997}.  This EOS contains
the single parameter, $\epsilon_c$, which is the transition energy density
between the low-density EOS with $P=0$ and the high-density EOS with
$P=\epsilon-\epsilon_c$.  The value of $\mathcal{C}$ for the maximum mass
case is 1.35 kHz.

In Fig.\
\ref{fig:MRrot} we also displayed  the correspondence between
the stellar configurations of the same baryon number. The line
${\rm S}_{\rm max} \to {\rm K}^0_{\rm max}$ separates "supramassive"
configurations from the "normal" ones, which can be reached by
spinning up  a non-rotating star. The maximum rotational
frequency for the "normal" sequences (reached at point
${\rm K}^0_{\rm max}$) has been discussed by \citet{CST1994a,CST1994b}
for polytropic and realistic EOSs ( their
Tables 3 and 7 respectively). It should be noted that this
value cannot be estimated using  our formula for $f_{\rm K}(M)$,
 Eq.\ (\ref{eq:fmax.form}), because  our formula  is valid
 within a restricted mass range
 $0.5 M_\odot <M < 0.9M_{\rm max}^{\rm stat}$.

Formula (6) connects configurations of the same gravitational mass $M$.
For example, it connects  $M=1.4\;M_\odot$ non-rotating star (filled circle
- ${\rm S}_{1.4}$)  and Keplerian configuration
(filled square - ${\rm K}_{1.4}$),
in Fig.\;5. They are joined by a solid horizontal line. At a fixed
baryon mass, $M_{\rm b}$, $M$ increases with increasing rotation
frequency. Dot line connecting
filled circle (${\rm S}_{1.4}$) and an open square near ${\rm K}_{1.4}$
contains configurations  with fixed $M_{\rm b}$, equal to that of a non-rotating
star with $1.4~M_\odot$. Deviation of solid line from the dot one visualizes
the rotational increase  of $M$ at a fixed $M_{\rm b}$.  For a
$1.4~M_\odot$ star, the fractional increase  is equal to
1.7\%, and for the maximum static mass  (dot line ${\rm S}_{\max}
\to {\rm K^0}_{\rm max}$) it reaches 3.8\%.

\section{Relation between  $R(M)$ and $R_{\rm K}(M)$}
\label{sect:Rmax-R}
\begin{figure}[h]
\centering
\resizebox{3.5in}{!}{\includegraphics[angle=0]{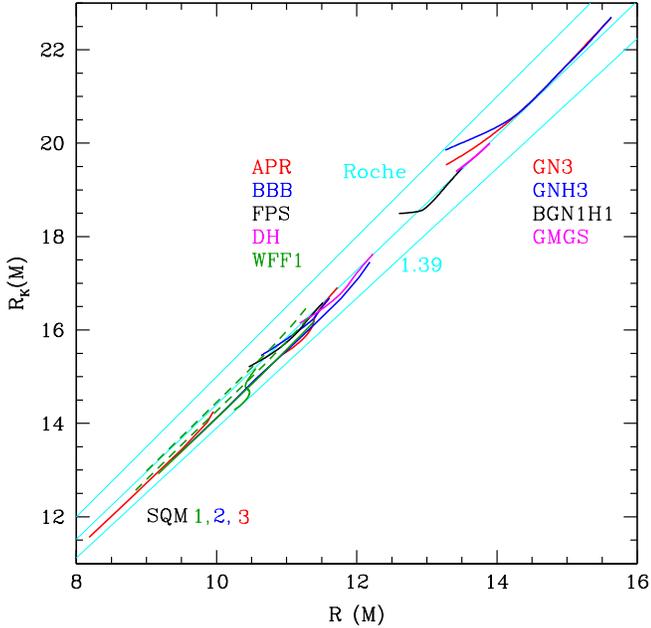}}
\caption{(Color online) Equatorial circumferential radius of the
Keplerian configuration, $R_{\rm K}(M)$, vs. circumferential radius
of the static configuration $R$ of the same gravitational mass $M$,
for $0.5M_\odot<M<0.9 M_{\rm max}^{\rm stat}$, for neutron stars
(solid lines) and quark stars with crust (dash lines). Three
straight cyan lines correspond: upper line to  $a_{\rm
Roche}=1.5$, middle line to $a=a_{\rm NS}=1.44$, and bottom line
$a=1.39$. {\it Neutron stars}: Color of a curve for a given EOS
coincides with that of the EOS label (APR,\ldots,WFF1).
{\it Quark stars}: nearly straight green, blue, and  red solid lines,
 located in the lower bundle,  correspond to the SQM1,
SQM2, and SQM3 EOSs of Table \ref{tab:EOSs-SQM} with a maximum solid
crust. Green dash lines in this bundle describe results obtained for
the SQM1 EOS of quark core and low-mass crusts:  $0.3\;M_{\rm
cr,max}$ (middle green line) and $0.06\;M_{\rm cr,max}$ (upper green
line).}
 \label{fig:R_K.R}
\end{figure}

Let us consider a family (sequence) of stationary
configurations rotating stably at a frequency $f$. They form a
curve in the $M-R_{\rm eq}$ plane (see examples in
Fig.\;\ref{fig:MRrot}). The curve is bound at
$R_{\rm eq}=R_{\rm min}(f)$
by the axisymmetric instability, implying star collapse into
a Kerr black hole. The largest circumferential radius is
reached for  the Keplerian configuration, $R_{\rm max}(f)=
R_{\rm K}(M)$. \cite{Bejger1122Hz} have shown that $R_{\rm
max}(f)$ is (within 2\%) equal to the radius of an orbit of a
point particle moving in the Schwarzschild space-time around a
point (or a spherical)  mass $M$. This implies
\begin{equation}
R_{\rm max}(f)\approx \left( {GM\over 4\pi^2 f^2}\right)^{1/3}~.
\label{eq:Rmax-f}
\end{equation}
For convenience we introduce a frequency $f_0$,
\begin{equation}
f_0={1\over 2\pi}\sqrt{GM_\odot\over (10~{\rm km})^3}=
1.8335~{\rm kHz}~.
\label{eq:f_0-def}
\end{equation}
Validity of the empirical formula, Eq.\ (\ref{eq:fmax.form}),
suggests then an approximate  proportionality
\begin{equation}
R_{\rm K}(M)\approx a\; R(M)~,~~a=\left( {C\over f_0}\right)^{-2/3}.
\label{eq:R_K-R}
\end{equation}
For the extreme relativistic Roche model,  $R_{\rm K}(M)$ is
{\it strictly} proportional to $R(M)$, and $a_{\rm Roche}=1.5$, Eq.\
(\ref{eq:R_K.Roche}). For  neutron stars and quark stars with crusts,
with  masses within $0.5M_\odot<M<0.9 M_{\rm max}^{\rm stat}$,  the
proportionality holds within a few percent, as shown on Fig.\ \ref{fig:R_K.R}.
However, the best-fit proportionality factors are smaller
than 1.5 of the Roche model, $a_{\rm NS}\approx a_{\rm QS}
\approx 1.44$.

The dependence of $a_{\rm QS}$ on the crust mass, $M_{\rm cr}$,  is
very weak. This can be explained via the effects of $M_{\rm cr}$ on
$R(M)$ and $f_{\rm K}(M)$. These effects oppose themselves: at fixed
$M$, $R(M)$ increases, and $f_{\rm K}(M)$ decreases, with an
increasing $M_{\rm cr}$. The cancellation of both effects results in
an only slight decrease of $a_{\rm QS}$ with increase of $M_{\rm
cr}$ (see Fig.\ \ref{fig:R_K.R}).

\cite{Lasota1996} derived an approximate relation between
equatorial radius of a maximally rotating configuration,
$R^{\rm rot}_{f_{\rm max}}$ and the radius of non-rotating
neutron star with maximum allowable mass, $R^{\rm stat}_{M_{\rm max}}$.
Maximally rotating configuration, stable both with respect to mass shedding
{\it and} axisymmetric perturbations, is actually very close
to that with largest mass, $M^{\rm rot}_{\rm max}$ (in Fig.\ \ref{fig:MRrot}
they are indistinguishable). The approximate proportionality
found by \cite{Lasota1996} for neutron stars  is
$R^{\rm rot}_{f_{\rm max}}\approx 1.32~\; R^{\rm stat}_{M_{\rm max}}~.$
This relation connects two extremal configurations. They have
different masses,  related by $M_{\rm max}^{\rm rot}\approx 1.18
M_{\rm max}^{\rm stat}$ \citep{Lasota1996}. In contrast, Eq.\
(\ref{eq:R_K-R}) connects normal configurations of neutron stars
{\it and}  quarks stars   with same gravitational mass
and holds for $0.5M_\odot<M<0.9M_{\rm
max}^{\rm stat}$.
\section{Discussion and conclusions}
\label{sect:discuss.concl}
We have tested empirical formula for Keplerian (mass shedding) frequency
of neutron star of mass $M$, proposed by \citet{LP2004}. Using
numerical results of precise 2-D calculations, performed for ten
representative realistic EOSs of dense matter based on different
dense matter models, we find prefactor $C_{\rm NS}=1.08~$kHz,
slightly higher than 1.04~kHz proposed by \citet{LP2004}. With our
prefactor, the formula is quite precise for $0.5M_\odot<M<0.9M_{\rm
max}$ (typically within 2\%, maximum deviation occurring for highest
$M$ not exceeding 6\%). Quark stars can reach larger $f_{\rm K}(M)$
than neutron stars. With a maximum crust on quark stars, we get
$C_{\rm QS}=1.15~$kHz. The value of $C_{\rm QS}$ does not depend
significantly on the crust mass, and can be used also for bare quark
stars.  We notice that both $C_{\rm NS}$ and $C_{\rm QS}$ are
significantly larger than for the relativistic Roche model, $C_{\rm
Roche}=1.00~$kHz \citep{STW1983}.

Using an approximate but quite precise
 Schwarzschild-like formula, relating $M$,
$R_{\rm K}$, and $f_{\rm K}$ \citep{Bejger1122Hz}, we show that to a
very good approximation the mass-shedding radius at a given $M$ is
proportional to the static radius $R(M)$, provided
$0.5\;M_\odot<M<0.9\;M_{\rm max}^{\rm stat}$. For neutron stars and
quark stars  we obtain the best-fit proportionality factor $a_{\rm
NS}\approx a_{\rm QS}\approx  1.44$.
 These proportionality factors are  smaller than the exact
factor 1.5 obtained for the relativistic Roche model.

Concluding, we derived a set of empirical formulae, expressing Keplerian
frequency and equatorial radius of Keplerian configuration in terms of the
mass and radius of normal configuration of the same mass. These formulae can
be used for masses $0.5M_\odot<M<0.9M_{\rm max}$. The formulae are approximate
but quite precise, and therefore might be  useful for constraining
the EOS of dense matter by the observations of pulsars.
\acknowledgements{This work was partially supported by the Polish
MNiSW grant no. N20300632/0450 and by the US DOE grant DE-AC02-87ER40317.
MB was partially supported
by Marie Curie Fellowship no.
ERG-2007-224793 within the 7th European Community Framework Programme.}

\end{document}